\documentclass[aps,superscriptaddress,preprint,11pt]{revtex4}

\usepackage{amsmath}
\usepackage{graphics}
\usepackage{graphicx}
\usepackage[mathcal]{eucal}

\begin{document}



\preprint{COLO-HEP-527, UCI-TR-2007-21, FERMILAB-PUB-07-108-T}

\title{CKM and Tri-bimaximal MNS Matrices in a $SU(5) \times { }^{(d)}T$ Model}


\author{Mu-Chun Chen}
\email[]{mcchen@fnal.gov}
\affiliation{Department of Physics \& Astronomy\\
University of California, Irvine, CA 92697, USA}
\affiliation{Theoretical Physics Department, Fermilab, Batavia, IL 60510-0500, USA}
\author{K.T. Mahanthappa}
\email[]{ktm@pizero.colorado.edu}
\affiliation{Department of Physics, University of Colorado at Boulder, Boulder, CO 80309-0390, USA}


\date{\today}

\begin{abstract}
We propose a model based on $SU(5) \times { }^{(d)}T$ which successfully gives rise to near tri-bimaximal leptonic mixing as well as realistic CKM matrix elements for the quarks. The Georgi-Jarlskog relations for three generations are also obtained. Due to the ${ }^{(d)}T$ transformation property of the matter fields, the $b$-quark mass can be generated only when the ${ }^{(d)}T$ symmetry is broken, giving a dynamical origin for the hierarchy between $m_{b}$ and $m_{t}$.  There are only nine operators allowed in the Yukawa sector up to at least mass dimension seven due to an additional $Z_{12} \times Z_{12}^{\prime}$ symmetry, which also forbids, up to some high orders, operators that lead to proton decay. 
The resulting model has a total of nine parameters in the charged fermion and neutrino sectors, and hence is very predictive. 
In addition to the prediction for $\theta_{13} \simeq \theta_{c}/3\sqrt{2}$, the model gives rise to a sum rule, $\tan^{2}\theta_{\odot} \simeq \tan^{2} \theta_{\odot, \mathrm{TBM}} - \frac{1}{2} \theta_{c}  \cos\beta$,  which is a consequence of the Georgi-Jarlskog relations in the quark sector. 
This deviation could account for the difference between the experimental best fit value for the solar mixing angle and the value predicted by the tri-bimaximal mixing matrix.
\end{abstract}


\maketitle

\section{Introduction}

The measurements of neutrino oscillation parameters have entered a precision era. The global fit to current data from neutrino oscillation experiments give the following best fit values and $2\sigma$ limits for the mixing parameters~\cite{Maltoni:2004ei},
\begin{equation}
\sin^{2} \theta_{12} = 0.30 \; (0.25 - 0.34), \quad
\sin^{2} \theta_{23} = 0.5 \; (0.38 - 0.64), \quad
\sin^{2} \theta_{13} = 0 \;  (< 0.028) \; .
\end{equation}
These values for the mixing parameters are very close to the values arising from the so-called ``tri-bimaximal'' mixing (TBM) matrix~\cite{Harrison:2002er},
\begin{equation}
U_{\mathrm{TBM}} = \left(\begin{array}{ccc}
\sqrt{2/3} & 1/\sqrt{3} & 0\\
-\sqrt{1/6} & 1/\sqrt{3} & -1/\sqrt{2}\\
-\sqrt{1/6} & 1/\sqrt{3} & 1/\sqrt{2}
\end{array}\right) \; , \label{eq:tri-bi}
\end{equation}
which predicts $\sin^{2}\theta_{\mathrm{atm, \, TBM}} = 1/2$ and $\sin\theta_{13, \mathrm{TBM}} = 0$. In addition, it predicts $\sin^{2}\theta_{\odot, \mathrm{TBM}} = 1/3$ for the solar mixing angle. Even though the predicted $\theta_{\odot, \mathrm{TBM}}$ is currently still  allowed by the experimental data at $2\sigma$, as it is very close to the upper bound at the $2\sigma$ limit, it  may be ruled out once more precise measurements are made in the  upcoming experiments.  

The finite group $A_{4}$, which describes the even permutations of four objects and possesses four in-equivalent representations, $1$, $1^{\prime}$, $1^{\prime \prime}$ and $3$, has been utilized as a family symmetry~\cite{Ma:2001dn}. 
It was pointed out that the tri-bimaximal mixing matrix can arise when the $A_{4}$ family symmetry  is imposed in the lepton sector~\cite{Ma:2004zv}.  However, due to its lack of doublet representations, CKM matrix is an identity in most $A_{4}$ models. In addition, to explain the mass hierarchy among the charged fermions, one needs to resort to additional symmetry.  It is hence not easy to implement $A_{4}$ as a family symmetry for both quarks and 
leptons~\cite{Ma:2006sk}.  

In this letter, we consider a different finite group, the double tetrahedral group, ${ }^{(d)}T$, which is a double covering of $A_{4}$. (For a classification of all finite groups up to order $32$ that can potentially be a family symmetry, see~\cite{Frampton:1994rk}). Because it has the same four in-equivalent representations as in $A_{4}$, the tri-bimaximal mixing pattern can be reproduced. In addition,  ${ }^{(d)}T$ has three in-equivalent doublets, $2$, $2^{\prime}$, and $2^{\prime\prime}$, which can be utilized to give the $2+1$ representation assignments for the quarks~\cite{Carr:2007qw}. 
In the context of SU(2) flavor group, this assignment has been known to give realistic quark mixing matrix and mass hierarchy~\cite{Chen:2000fp}. Utilizing ${}^{(d)}T$ as a family symmetry for both quarks and leptons has been considered before in a SU(5) model~\cite{Aranda:1999kc} and in 
a non-unified model~\cite{Feruglio:2007uu}. In Ref.~\cite{Aranda:1999kc}, both quarks and leptons (including the neutrinos) have $2\oplus 1$ representation assignments under ${}^{(d)}T$, and the prediction for the solar mixing angle is $\sim 10^{-3}$, which is in the region of small mixing angle solution that has been ruled out by SNO and KamLAND. A recent attempt in \cite{Feruglio:2007uu} generalizes the ${}^{(d)}T$ to the quark sector while maintaining near TBM pattern. However,  in order to explain the mass hierarchy, the model has to resort to an additional $U(1)$ symmetry. Furthermore, a large number of operators are present in this model, making it less predictive.  Here we consider  an SU(5) model combined with ${}^{(d)}T$ symmetry, which successfully accommodates the mass hierarchy as well as the mixing matrices in both quark and lepton sectors. With an additional $Z_{12}\times Z_{12}^{\prime}$ symmetry, only ``good'' operators are allowed up to at least dimension seven, making the model very predictive.  In addition, the mass hierarchy is naturally explained without having hierarchy in the vacuum expectation values (VEV's) of the scalar fields, the reason being that the mass operators for the lighter generation are allowed to appear only at higher order compared to those for the heavy generation. Thus we have a dynamical explanation for the mass hierarchy. 

\section{The Model}

In SU(5), all matter fields are unified into a $10(Q,\, u^{c}, \, e^{c})_{L}$  and a $\overline{5}(d^{c}, \ell)_{L} $ dimensional representations. 
The three generations of $\overline{5}$ are assigned into a triplet of ${ }^{(d)}T$, in order to generate the tri-bimaximal mixing pattern in the lepton sector, and it is denoted by $\overline{F}$. On the other hand, to obtain realistic quark sector, the third generation of the 10-dim representation transforms as a singlet, so that the top quark mass is allowed by the family symmetry, while the first and the second generations form a doublet of ${ }^{(d)}T$. These 10-dim representations are denoted by, respectively, $T_{3}$ and $T_{a}$, where $a = 1,2$. The Yukawa interactions are mediated by a 5-dim Higgs, $H_{5}$, a $\overline{5}$-dim Higgs, $H_{\overline{5}}^{\prime}$, as well as a 45-dim Higgs, $\Delta_{45}$, which is required for the Georgi-Jarlskog relations. We have summarized these quantum number assignment in Table~\ref{tbl:charge}. It is to be noted that $H_{5}$ and $H_{\overline{5}}^{\prime}$ are not conjugate of each other as they have different $Z_{12}$ and $Z_{12}^{\prime}$ charges. 
 \begin{table}
\begin{tabular}{|c|ccc|ccc|cccccc|cc|}\hline
& $T_{3}$ & $T_{a}$ & $\overline{F}$ & $H_{5}$ & $H_{\overline{5}}^{\prime}$ & $\Delta_{45}$ & $\phi$ & $\phi^{\prime}$ & $\psi$ & $\psi^{\prime}$ & $\zeta$ & $N$ & $\xi$ & $\eta$  \\ [0.3em] \hline\hline
SU(5) & 10 & 10 & $\overline{5}$ & 5 &  $\overline{5}$ & 45 & 1 & 1 & 1 & 1& 1 & 1 & 1 & 1\\ \hline
${ }^{(d)}T$ & 1 & $2$ & 3 & 1 & 1 & $1^{\prime}$ & 3 & 3 & $2^{\prime}$ & $2$ & $1^{\prime\prime}$ & $1^{\prime}$ & 3 & 1 \\ [0.2em] \hline
$Z_{12}$ & $\omega^{5}$ & $\omega^{2}$ & $\omega^{5}$ & $\omega^{2}$ & $\omega^{2}$ & $\omega^{5}$ & $\omega^{3}$ & $\omega^{2}$ & $\omega^{6}$ & $\omega^{9}$ & $\omega^{9}$ 
& $\omega^{3}$ & $\omega^{10}$ & $\omega^{10}$ \\ [0.2em] \hline
$Z_{12}^{\prime}$ & $\omega$ & $\omega^{4}$ & $\omega^{8}$ & $\omega^{10}$ & $\omega^{10}$ & $\omega^{3}$ & $\omega^{3}$ & $\omega^{6}$ & $\omega^{7}$ & $\omega^{8}$ & $\omega^{2}$ & $\omega^{11}$ & 1 & $1$ 
\\ \hline   
\end{tabular}
\vspace{0.1in}
\caption{Charge assignments. Here the parameter $\omega = e^{i\pi/6}$.}  
\label{tbl:charge}
\end{table}

The Lagrangian of the model is given as follows,
\begin{eqnarray}
\mathcal{L}_{\mathrm{Yuk}} & = &  \mathcal{L}_{\mathrm{TT}} + \mathcal{L}_{\mathrm{TF}} + \mathcal{L}_{\mathrm{FF}} \\
\mathcal{L}_{\mathrm{TT}} & = & y_{t} H_{5} T_{3}T_{3} + \frac{1}{\Lambda^{2}} y_{ts} H_{5} T_{3} T_{a} \psi \zeta + 
\frac{1}{\Lambda^{2}} y_{c} H_{5} T_{a} T_{a} \phi^{2} + \frac{1}{\Lambda^{3}} y_{u} H_{5} T_{a} T_{a} \phi^{\prime 3}   \\  
\mathcal{L}_{\mathrm{TF}} & = & \frac{1}{ \Lambda^{2}}  y_{b} H_{\overline{5}}^{\prime} \overline{F} T_{3} \phi \zeta + \frac{1}{\Lambda^{3}} \biggl[ y_{s} \Delta_{45} \overline{F} T_{a} \phi \psi N  + 
 y_{d} H_{\overline{5}}^{\prime} \overline{F} T_{a} \phi^{2} \psi^{\prime}  \biggr]\\
\mathcal{L}_{\mathrm{FF}} & = & \frac{1}{M_{x}\Lambda} \biggl[\lambda_{1} H_{5} H_{5} \overline{F} \, \overline{F} \xi +  \lambda_{2} H_{5} H_{5} \overline{F} \, \overline{F} \eta\biggr] \; ,
 \end{eqnarray}
 where $M_{x}$ is the cutoff scale at which the lepton number violation operator $HH\overline{F}\, \overline{F}$ is generated, while $\Lambda$ is the cutoff scale, above which the ${}^{(d)}T$ symmetry is exact.  The parameters $y$'s and $\lambda$'s are the coupling constants. 
The vacuum expectation values (VEV's) of various SU(5) singlet scalar fields are,
\begin{eqnarray}
{ }^{(d)}T \longrightarrow G_{\mathrm{TST^{2}}}: & ~~~
\bigl< \xi \bigr> = \xi_{0} \Lambda\left(\begin{array}{c}
1 \\ 1 \\ 1\end{array}\right), \quad 
\bigl< \phi^{\prime} \bigr> = \phi_{0}^{\prime} \Lambda\left(\begin{array}{c}
1 \\ 1 \\ 1\end{array}\right), \quad
 \\
{ }^{(d)}T \longrightarrow G_{\mathrm{T}}: & ~~~
\bigl< \phi \bigr> = \phi_{0} \Lambda\left(\begin{array}{c}
1 \\ 0 \\ 0 \end{array}\right), \quad 
 \bigl< \psi \bigr> = \psi_{0} \Lambda\left(\begin{array}{c}
1 \\ 0\end{array}\right) \\
{ }^{(d)}T \longrightarrow \mbox{nothing}: & ~~~
\bigl< \psi^{\prime} \bigr> = \psi^{\prime}_{0} \Lambda\left(\begin{array}{c}
1 \\  1 \end{array}\right)~~\\
{ }^{(d)}T \longrightarrow G_{\mathrm{S}}: & ~~~
\bigl< \zeta \bigr> = \zeta_{0} \Lambda, \quad \bigl<N\bigr> =N_{0} \Lambda \\
{ }^{(d)}T- \mbox{invariant}: & ~~~
\bigl< \eta \bigr> = u \Lambda
\end{eqnarray}
where $G_{\mathrm{TST^{2}}}$ denotes the subgroup generated by the elements $TST^{2}$, which in the triplet representation is given by~\cite{Feruglio:2007uu},
\begin{equation}
TST^{2} = \frac{1}{3}\left( \begin{array}{ccc}
-1 & 2 & 2\\
2 & -1 & 2\\
2 & 2 & -1
\end{array}\right) \; , 
\end{equation}
while $G_{\mathrm{T}}$ and $G_{\mathrm{S}}$ denote subgroup generated by the elements $T$ and $S$, respectively. (Our notation is the same as in Ref.~\cite{Feruglio:2007uu}.) The details concerning vacuum alignment of these VEV's will be presented in a future publication.

We have summarized the remaining operators in the charged fermion sectors that are otherwise allowed by the $SU(5) \times { }^{(d)}T$ symmetry in Table~\ref{tbl:operators}. 
By imposing an additional $Z_{12} \times Z_{12}^{\prime}$ symmetry, under which the transformation properties of various fields are summarized in Table~\ref{tbl:charge}, the above  Lagrangian is the most general one. Here the operators that couple to $H_{5} T_{3} T_{3}$ are not shown in the above Lagrangian as their contributions can be absorbed into a redefinition of the coupling constant $y_{t}$. In addition, we neglect the operator $H_{\overline{5}}^{\prime} \overline{F} T_{3} \zeta \psi \psi^{\prime}$  in $\mathcal{L}_{TF}$ since its contribution is negligible. Also not shown are those that contribute to $\mathcal{L}_{\mathrm{FF}}$ which can be absorbed into a redefinition of the parameter $u$ and $\phi_{0}$. Note that in principle, viable phenomenology may still be obtained when more operators are allowed. The additional discrete symmetry that is needed in that case would be smaller. Nevertheless, more Yukawa coupling constants will be present and the model would not be as predictive. The $Z_{12} \times Z_{12}^{\prime}$ symmetry also forbids proton and other nucleon decay operators to very high orders; it is likely this symmetry might be linked to orbifold compactification in extra dimensions.  
Note that, the $Z_{12}\times Z_{12}^{\prime}$ symmetry also separates the neutrino and charged fermion sectors, so that the neutrinos only couple to the $G_{\mathrm{TST^{2}}}$ breaking sector.  
Furthermore, it allows the 45-dim Higgs, $\Delta_{45}$, to appear only in the operator shown above, and thus is crucial for obtaining the Georgi-Jarlskog (GJ) relations.  

\begin{table}[t]
\begin{tabular}{|c|l|}
\hline
$H_{5} T_{3} T_{a}$ ~~ & ~~ $\psi^{\prime}, \; \psi$\\
& ~~$\psi \phi, \; \psi \phi^{\prime}, \; \psi^{\prime} \phi, \; \psi^{\prime} \phi^{\prime}, \; \psi^{\prime}\zeta, \; \psi^{\prime}N, \; \psi N$ \\
& ~~$\psi^{3}, \; \psi \psi^{\prime 2}, \; \psi\phi^{2}, \; \psi\phi^{\prime 2}, \; \psi\phi \zeta, \; \psi\phi^{\prime} \zeta, \; 
\psi^{\prime 3}, \; \psi^{\prime} \psi^{2}, \; \psi^{\prime} \phi^{2}, \; \psi^{\prime} \phi^{\prime 2}, \; \psi^{\prime}\phi \zeta, \; \psi^{\prime}\phi^{\prime} \zeta, \;$ ~~\\ 
& ~~~~~~$\psi\phi N, \psi \phi^{\prime}N, \; \psi^{\prime} \phi N, \psi^{\prime} \phi^{\prime}N$ ~~\\
\hline
& ~~$\psi \xi, \; \psi^{\prime} \xi, \; \psi \xi^{2}, \; \psi\xi \phi, \; \psi\xi\phi^{\prime}, \; \psi\xi\zeta, \; 
\psi^{\prime} \xi^{2} \; \psi^{\prime} \xi \phi, \; \psi^{\prime} \xi \phi^{\prime}, \; 
\psi^{\prime} \xi \zeta, \; \psi \xi N, \; \psi^{\prime} \xi N, \; \psi^{\prime} \eta, \; \psi \phi \eta, \; 
\psi \phi^{\prime} \eta, \; \psi \xi \eta,$ \; \\
&~~~~~~
$\psi^{\prime} \phi \eta, \; \psi^{\prime} \phi^{\prime} \eta, \; 
 \psi^{\prime} \xi \eta, \; \psi \eta, \; 
  \psi \phi \eta, \; 
 \psi\phi^{\prime} \eta, \;
 \psi^{\prime} \phi \eta, \; \psi^{\prime} \phi^{\prime} \eta, \; 
 \psi \phi \eta, \; \psi \phi^{\prime} \eta, \; 
 \psi^{\prime} \phi \eta, \; \psi^{\prime} \phi^{\prime} \eta$
\\ \hline
$H_{5} T_{a} T_{a}$~~ & ~~ $\phi$, \; $\phi^{\prime}$ \\
& ~~$\phi^{\prime 2}, \; \psi^{2}, \; \psi^{\prime 2}, \; \phi \phi^{\prime}, \; \psi \psi^{\prime}$\\
& ~~$\phi^{3}, \; \phi^{2} \zeta, \; \phi \zeta^{2},  \; \phi^{\prime 2} \zeta, \; \phi^{\prime} \zeta^{2}, \phi\phi^{\prime}\zeta, \; \phi\phi^{\prime2}, \; \phi^{\prime}\phi^{2}, \; 
\phi N^{2}, \; \phi^{\prime} N^{2}, \; \phi^{\prime 2} N, \; \phi\phi^{\prime} N, \; \phi N \zeta, \; \phi^{\prime} N \zeta$ ~~\\
\hline
&~~ $\xi, \; \xi^{2}, \; \xi\zeta, \; \xi N, \; \xi\eta, \; \xi^{2}, \; 
\xi\phi, \; \xi\phi^{\prime}, \; \xi^{3}, \; 
\xi^{2} \zeta, \; \xi^{2} \eta, \xi^{2} \zeta, \; \xi N \zeta, \; \xi N \eta, \; \xi\zeta \eta, \; \xi\phi^{2}, \; \xi\phi^{\prime 2}, \; 
\xi\phi\phi^{\prime},$ \;\\
& ~~~~~~ $\xi^{2} \phi, \; \xi^{2} \phi^{\prime}, \; \xi\phi N, \; \xi \phi \eta, \; \xi \phi\zeta, \; \xi\phi^{\prime} N, \; \xi\phi^{\prime} \eta, \; \xi\phi^{\prime}\zeta, \;
\phi^{2}\eta, \; \phi \eta^{2}, \; \phi\eta N, \; \phi\eta \zeta, \; 
\phi^{\prime}\eta^{2}, \; \phi^{\prime} \eta N,$ \\
&~~~~~~$\phi^{\prime} \eta \zeta, \phi \eta, \phi^{\prime} \eta, \; \xi N^{2}, \; \xi\eta^{2}, \; \xi\zeta^{2}$
\\ \hline
$H_{\overline{5}}^{\prime} \overline{F} T_{3}$ ~~ & ~~$\phi, \; \phi^{\prime}$  \\
 & ~~$\psi^{2}, \; \phi^{2}, \; \phi^{\prime 2}, \; \phi^{\prime}\phi, \; \psi^{\prime 2}, \; \psi\psi^{\prime}, \; \phi^{\prime}\zeta, \; \phi^{\prime} N, \; \phi N$\\
 & ~~$\phi^{3}, \; \phi^{\prime 3}, \; \phi^{2} \phi^{\prime}, \; \phi \phi^{\prime 2}, \; \phi \zeta^{2}, \; \phi^{\prime} \zeta^{2}, \phi \psi^{2}, \; \phi^{\prime} \psi^{\prime 2}, \; \zeta\psi^{2}, \; 
 \zeta\psi^{\prime2}, \; \phi^{\prime} \psi^{2}, \; \phi \psi^{2}, $~~\\
 & ~~~~~~ $ \phi N^{2}, \; \phi^{\prime} N^{2}, \; \phi N \zeta, \; \phi^{\prime} N \zeta, \; N\psi^{2}, \; \zeta \psi^{2}, \; \zeta \psi\psi^{\prime}, \; N\psi \psi^{\prime}$\\  \hline
 & ~~$ \xi, \; \xi^{2}, \; \xi N, \; \xi\zeta, \; \xi\eta, \; \xi\phi, \; \xi\phi^{\prime}, \; \xi^{3}, \; \xi^{2} N, \; \xi^{2} \zeta, \; \xi^{2} \eta, \; 
 \xi^{2}\phi, \; \xi^{2} \phi^{\prime}, \; \xi\phi^{2}, \;$\\
 &~~~~~~$ \xi \phi^{\prime 2}, \; 
 \xi\phi\phi^{\prime}, \; \xi\phi N, \; \xi\phi \zeta, \; \xi\phi \eta, \; \xi\phi^{\prime} N, \; \xi\phi^{\prime} \zeta, \; \xi\phi^{\prime} \eta, \; \phi^{\prime} \eta, \; \phi \eta^{2}, \; \phi\eta N, \; \phi \eta\zeta, \; \phi^{\prime} \eta^{2}, \; 
 \phi^{\prime} \eta N, \; \phi^{\prime} \eta \zeta, \; \eta \psi^{2}, $\\
 &~~~~~~$\; \eta \psi^{\prime 2}, \; \phi\eta, \;  \phi\eta N, \; \phi\eta \zeta, \; 
 \phi^{\prime} \eta^{2}, \; \phi^{\prime} \eta N, \; \eta \psi\psi^{\prime} $
 \\
 \hline
$H_{\overline{5}}^{\prime} \overline{F}T_{a}$ ~~  & ~~ $\psi, \; \psi^{\prime}$\\
& ~~$\psi \phi^{\prime}, \; \psi^{\prime} \phi, \; \psi^{\prime} \phi^{\prime}, \; \phi\psi$\\
& ~~$\psi \phi^{2}, \;  \psi \phi \zeta, \; \psi^{\prime} \phi \zeta, \; \psi\phi^{\prime 2}, \; \psi^{\prime} \phi^{\prime 2}, \; \psi \phi \phi^{\prime},
 \; \psi^{\prime} \phi \phi^{\prime}, \; \psi\phi^{\prime} \zeta, \;  \psi^{\prime}\phi^{\prime} \zeta, \; \psi\phi N, \; \psi^{\prime} \phi N, \; \psi\phi^{\prime} N, \; \psi^{\prime} \phi^{\prime} N ~~$\\ \hline
 &~~$\psi\xi, \; \psi^{\prime}\xi, \; \psi\xi^{2}, \; \psi^{\prime}\xi^{2}, \; \psi\xi\phi, \; \psi\xi\phi^{\prime}, \; \psi^{\prime}\xi\phi, \; \psi^{\prime}\xi\phi^{\prime}, \; $\\
 &~~~~~~$\psi\xi N, \; \psi \xi\eta, \psi \xi\zeta, \; \psi^{\prime}\xi\zeta, \; \psi^{\prime}\xi \eta, \; \psi^{\prime}\xi N, \; \psi \phi\eta, \; \psi^{\prime}\phi\eta, \; 
 \psi^{\prime}\phi^{\prime} \eta, \; \psi \phi^{\prime}\eta, \; \psi^{\prime}\phi^{\prime}\eta, \; \psi\phi \eta, \; \psi^{\prime} \phi \eta$
\\ \hline
\end{tabular}
\vspace{0.1in}
\caption{Additional operators that are allowed by the $SU(5) \times { }^{(d)}T$ symmetry up to dim-7. For each operator shown above, there is a corresponding one with $H_{5}, \, H_{\overline{5}}^{\prime} \leftrightarrow \Delta_{45}$.}
\label{tbl:operators}
\end{table}

The interactions in $\mathcal{L}_{\nu}$ give the following neutrino mass matrix~\cite{Ma:2001dn}, which is invariant under $G_{\mathrm{TST^{2}}}$~\cite{Feruglio:2007uu}, 
\begin{equation}
M_{\nu} = 
\frac{\lambda v^{2}}{M_{x}} \left(\begin{array}{ccc}
2 \xi_{0} + u & -\xi_{0} & -\xi_{0} \\
-\xi_{0} & 2 \xi_{0} & u - \xi_{0} \\
-\xi_{0} & u-\xi_{0} & 2 \xi_{0}\end{array}\right) \; ,
\end{equation} 
and we have absorbed the Yukawa coupling constants by rescaling the VEV's.  
This mass matrix $M_{\nu}$ is form diagonalizable, {\it i.e.} the orthogonal matrix that diagonzlizes it does not depend on the eigenvalues. Its diagonal form is,
\begin{equation}
V_{\nu}^{\mathrm{T}} M_{\nu} V_{\nu} = \mathrm{diag}(u+ 3\xi_{0}, \, u, \, -u + 3\xi_{0}) \frac{v^{2}_{u}}{M_{x}} \; ,
\end{equation}
where the diagonalization matrix $V_{\nu}$ is the tri-bimaximal mixing matrix, $V_{\nu} = U_{\mathrm{TBM}}$ given in Eq.~\ref{eq:tri-bi}. 
This tri-bimaximal mixing pattern and the mass eigenvalues in the neutrino sector are thus the same as in all previous analyses in models based on $A_{4}$ and ${}^{(d)}T$, which has been shown to be consistent with experimental data. 

The down type quark and charged lepton masses are generated by $\mathcal{L}_{TF}$. Because the renormalizable operator 
$H_{\overline{5}}^{\prime}  \overline{F} T_{3}$ is forbidden by the ${}^{(d)}T$ symmetry, the generation of $b$ quark mass requires the breaking of ${}^{(d)}T$, which naturally explains the hierarchy between $m_{t}$ and $m_{b}$. 
The $b$ quark mass, and thus the $\tau$ mass, is generated upon the breaking of ${ }^{(d)}T \rightarrow G_{\mathrm{T}}$ and ${ }^{(d)}T \rightarrow G_{\mathrm{S}}$. 
As $m_{b}$ and $m_{\tau}$ are generated by the same operator, $H_{\overline{5}}^{\prime} \overline{F} T_{3} \phi \zeta$, we obtain the successful $b-\tau$ unification relation. Upon the breaking of ${ }^{(d)}T \rightarrow G_{\mathrm{T}}$, the operator  $\Delta_{45} \overline{F} T_{a} \phi N$ contributes to the (22) element in $M_{d,\, e}$, and thus gives rise to $m_{s}$ and $m_{\mu}$. As this operator involves $\Delta_{45}$, the GJ relation for the second family, $m_{\mu} \simeq 3 m_{s}$ is obtained. If no further symmetry breaking takes place, the first generation masses, $m_{d}$ and $m_{e}$ vanishes. At this stage, the diagonalization mass matrix for the charged leptons (and down type quark) is identity, and hence the the tri-bimaximal mixing matrix is exact.  

To obtain the correct mass relation for the first generation, it inevitably calls for flavor mixing in the down quark sector, which then leads to corrections to the tri-bimaximal mixing pattern. The correction to the $\theta_{12}$ due to mixing in the charged lepton sector can account for the difference between $\sin^{2}\theta_{12} = 1/3$ in the tri-bimaximal mixing matrix and the experimentally observed best fit value, $\sin^{2} \theta_{12} = 0.3$. The GJ relation for the first family, $m_{d} \simeq 3 m_{e}$, is obtained due to the operator $H_{\overline{5}}^{\prime} \overline{F} T_{a} \phi^{2} \psi^{\prime}$, which further breaks the ${ }^{(d)}T$ symmetry down to nothing. 
The mass matrices for the down type quarks and charged leptons are thus given by,
 \begin{eqnarray}
M_{d} & = & \left( \begin{array}{ccccc}
0 & ~~ & (1+i) \phi_{0}\psi^{\prime}_{0} & ~~ & 0\\
-(1-i)\phi_{0}\psi^{\prime}_{0} & ~~ &   \psi_{0} N_{0} &  & 0\\
\phi_{0}\psi^{\prime}_{0} & ~~ & \phi_{0}\psi^{\prime}_{0} &  & \zeta_{0}
\end{array}\right) y_{b} v_{d} \phi_{0}, \label{eq:md} \\ 
M_{e} & = & \left( \begin{array}{ccccc}
0 & ~~ & -(1-i) \phi_{0}\psi^{\prime}_{0} & ~~ &  \phi_{0}\psi^{\prime}_{0}\\
(1+i)\phi_{0}\psi^{\prime}_{0} && -3 \psi_{0} N_{0} & & \phi_{0}\psi^{\prime}_{0}\\ 
0 & & 0 & & \zeta_{0}
\end{array}\right) y_{b} v_{d} \phi_{0} \label{eq:me}
\end{eqnarray}
where we have absorbed the coupling constants $y_{d}$ and $y_{s}$ by re-scaling the VEV's, $\phi_{0}$ and 
$\psi^{\prime}_{0}$.  Since the off diagonal elements in these mass matrices involve two VEV's, $\phi_{0} \psi_{0}^{\prime}$, they are naturally smaller compared to $\psi_{0}$, assuming the VEV's are naturally of the same order of magnitude. Besides explaining the mass hierarchy, it gives rise to the correct GJ relations in the first and the second families.  Furthermore, as $b$ is small, the corrections to $\theta_{12}$ and $\theta_{13}$ in the neutrino sector are under control. 
Note that there is no correction to $M_{d, \, e}$ given above at least to the order of dim-7. 
 
The up quark masses are generated by the following Yukawa interactions, $\mathcal{L}_{TT}$. When the ${ }^{(d)}T$ symmetry is exact, the only operator that is allowed is $H_{5} T_{3} T_{3}$, thus only top quark mass is generated, which naturally explains why the top mass is much larger than all other fermion masses. When $\bigl< \psi \bigr>$ breaks ${ }^{(d)}T$ down to $G_{\mathrm{T}}$, the mass $m_{c}$ and $V_{td}$ is generated by the operators, $H_{5}T_{3}T_{a} \phi \zeta$ and $H_{5} T_{a} T_{a} \phi^{2}$. The breaking of ${ }^{(d)}T \rightarrow G_{\mathrm{TST^{2}}}$ gives rise the up quark mass through the operator $H_{5} T_{a} T_{b} \phi^{\prime 3}$.  
These interactions give rise to the following mass matrix for the up type quarks,
\begin{equation}
M_{u} = \left( \begin{array}{ccccc}
i \phi_{0}^{\prime 3} & ~~ &  \frac{1-i}{2}  \phi_{0}^{\prime 3} & ~~ & 0\\
\frac{1-i}{2} \phi_{0}^{\prime 3} & & \phi_{0}^{\prime 3} +  (1-\frac{i}{2}) \phi_{0}^{2}  & & y^{\prime} \psi_{0}\zeta_{0}\\
0 & & y^{\prime} \psi_{0}\zeta_{0} & & 1
\end{array}\right) y_{t} v_{u} \; , \label{eq:mu}
\end{equation}
where we have absorbed $y_{c}/y_{t}$ and $y_{u}/y_{t}$ by re-scaling the VEV's of $\psi_{0}$ and $\phi_{0}^{\prime}$, and $y^{\prime} = y_{ts}/\sqrt{y_{c}y_{t}}$.

The mixing angel $\theta_{12}^{u}$ from the up type quark mass matrix given in Eq.~\ref{eq:mu} is related to  $m_{c}$ and $m_{u}$ as $\theta_{12}^{u} \simeq \sqrt{m_{u}/m_{c}}$, while the mixing angle $\theta_{12}^{d}$ arising from the down quark mass matrix $M_{d}$ given in Eq.~\ref{eq:md} is related to the ratio of $m_{d}$ and $m_{s}$ as $\theta_{12}^{d} \simeq \sqrt{m_{d}/m_{s}}$,  to the leading order.  The Cabibbo angle, $\theta_{c}$, is therefore given by $\theta_{c} \simeq \bigl| \sqrt{m_{d}/m_{s}} - e^{i\alpha} \sqrt{m_{u}/m_{c}} \, \bigr| \sim \sqrt{m_{d}/m_{s}}$, where the relative phase $\alpha$ depends upon the coupling constants. Even though $\theta^{d}_{12}$ is of the size of the Cabibbo angle, the corresponding mixing angle in the charged lepton sector, $\theta_{12}^{e}$, is much suppressed due to the GJ relations,
\begin{equation}
\theta_{12}^{e} \simeq \sqrt{\frac{m_{e}}{m_{\mu}}} \simeq \frac{1}{3} \sqrt{ \frac{m_{d}}{m_{s}}} \sim \frac{1}{3} \theta_{c} \; .
\end{equation} 
As a result, the correction to the tri-bimaximal mixing pattern due to the mixing in the charged lepton sector is small, and is given,  to the leading order, by,
\begin{equation}
\tan^{2}\theta_{\odot} \simeq \tan^{2} \theta_{\odot, \mathrm{TBM}} -  \frac{1}{2} \theta_{c} \cos\beta \; , 
\end{equation}
where the relative phase $\beta$ is determined by the strengths and phases of the VEV's, $\phi_{0}$ and $\psi_{0}^{\prime}$.  Such a relation was also found in a model based on Pati-Salam gauge group~\cite{King:2005bj} and is quite generic in models having Georgi-Jarlskog relations~\cite{Antusch:2005kw}. 
This deviation could account for the difference between the prediction of the TBM matrix, which gives $\tan^{2}\theta_{\odot, \mathrm{TBM}}=1/2$, and the experimental best fit value, $\tan^{2}\theta_{\odot,\mathrm{exp}}=0.429$, if $\cos\beta \simeq 2/3$ (with $\theta_{c} \simeq 0.22$). 
The off-diagonal matrix elements in $M_{e}$ also generate a non-zero value for the neutrino mixing angle $\theta_{13} \simeq \theta_{c}/3\sqrt{2} \sim 0.05$. We note that a more precise measurement of $\tan\theta_{\odot}$  will pin down the phase of $\phi_{0}\psi_{0}^{\prime}$, and thus the three leptonic CP phases, which may yield interesting consequences on leptogenesis~\cite{Chen:2007fv} and lepton flavor violating processes~\cite{Chen:2004xy}.  

\section{Numerical Results}

The observed quark masses respect the following relation, 
\begin{equation}
m_{u} : m_{c} : m_{t}  =  \epsilon_{u}^{2} : \epsilon_{u} : 1, \; \quad m_{d} : m_{s} : m_{b} =  \epsilon_{d}^{2} : \epsilon_{d} : 1 \; ,
\label{eq:ratio}
\end{equation}
where $\epsilon_{u} \simeq (1/200) = 0.005$ and $\epsilon_{d} \simeq (1/20) = 0.05$. 

In our model, the mass matrices for the down type quarks and charged leptons can be parametrized as,
\begin{equation}
\frac{M_{d}}{y_{b} v_{d} \phi_{0}\zeta_{0}} = \left( \begin{array}{ccccc}
0 & ~~ & (1+i) b & ~~ & 0\\
-(1-i) b & & c & & 0\\
b & &b & & 1
\end{array}\right),  \; \quad 
\frac{M_{e}}{y_{b} v_{d} \phi_{0}\zeta_{0}} = \left( \begin{array}{ccccc}
0 & ~~ &-(1-i) b & ~~ & b\\ 
(1+i) b & & -3c & & b\\
0 & & 0 & & 1
\end{array}\right) \; , 
\end{equation}
and with the choice of $b \equiv \phi_{0} \psi^{\prime}_{0} /\zeta_{0} = 0.00789$ and $c\equiv \psi_{0}N_{0}/\zeta_{0}  = - 0.0474$, the mass ratios for the down type quarks and for the charged leptons are given by,
\begin{eqnarray}
m_{d} : m_{s} : m_{b} & = & 0.00250 : 0.0499 : 1.00 \; , \\
m_{e} : m_{\mu} : m_{\tau} & = & 0.000870 : 0.143 : 1.00 \; .
\end{eqnarray} 
These predictions are consistent with the observed values given in Eq.~\ref{eq:ratio} and  are in good agreement with  the GJ relations. 
The overall scale factor is $y_{b} \phi_{0} \zeta_{0} \simeq m_{b}/m_{t} \simeq (0.011)$ at the GUT scale, assuming the top Yukawa coupling is 1.  
For the up type quarks, the mass matrix can be written as,
\begin{equation}
M_{u} = \left( \begin{array}{ccccc}
i g & ~~ &  \frac{1-i}{2}  g & ~~ & 0\\
\frac{1-i}{2} g & & g +  h  & & k\\
0 & & k & & 1
\end{array}\right) y_{t} v_{u} \; ,
\end{equation}
and with the choice of  $k\equiv y^{\prime}\psi_{0}\zeta_{0}= -0.032$, $h\equiv \psi_{0}^{2} = 0.0053$ and $g \equiv \phi_{0}^{\prime 3}=-2.25 \times 10^{-5}$, the ratio among the three up type quarks is given by,
\begin{equation}
m_{u} : m_{c} : m_{t} =  0.0000252 : 0.005 : 1.00 \; , \\
\end{equation}
which is consistent with the observed values shown in Eq.~\ref{eq:ratio}. The absolute values of the CKM matrix elements are given by,
\begin{equation}
|V_{\mathrm{CKM}}| = \left(\begin{array}{ccc}
0.976 & 0.219 & 0.00780\\
0.219 & 0.975 & 0.0400\\
0.00638 & 0.0403 & 0.999
\end{array}\right) \; .
\end{equation} 
Except for the element $V_{ub}$, which is slightly higher than the current experimental upper bound of $\sim 0.005$,  all other elements are in good agreement with current data. This discrepancy can be alleviated by allowing additional operators to be present in the model. It can also be improved by having complex parameters, with which realistic CP violation measures in the quark sector could also arise. We leave these possibilities for further investigation. The diagonalization matrix for the charged leptons is,
\begin{equation}
V_{e,L} = \left( \begin{array}{ccc}
-0.996 + 0.052i & -0.0517+0.0581i & ~(6.35-6.36i) \times 10^{-5} \\
0.0578+0.0520i & -0.995+0.0581i & 0.00108-0.0000636i \\
7.24\times 10^{-6}  ~& 0.00109 & 0.999
\end{array}\right). 
\end{equation}
This leads to small deviation to the tri-bimaximal mixing pattern as discussed above, leading to the following leptonic mixing matrix,
\begin{equation}
|U_{\mathrm{MNS}}| =|V_{e,L}^{\dagger} U_{\mathrm{TBM}}|= \left( \begin{array}{ccc}
0.838 & 0.545 & 0.0550\\
0.364 &  0.608 & 0.706 \\
0.409 & 0.578 & 0.706
\end{array}\right) \; ,
\end{equation}
which gives $\sin^{2}\theta_{\mathrm{atm}} = 1$, $\tan^{2}\theta_{\odot} = 0.424$ and $|U_{e3}| = 0.055$. 
Note that the total number of parameters in our model is seven in the charged fermion sectors and two in the neutrino sector.
 
\section{Conclusion}

In this letter, we have presented a grand unified model based on SU(5) combined with the double tetrahedral group, ${}^{(d)}T$, 
which successfully, for the first time, gives rise to near tri-bimaximal leptonic mixing as well as realistic CKM matrix elements for the quarks. Due to the presence of the  
$Z_{12} \times Z_{12}^{\prime}$ symmetry, only nine operators are allowed in the model, and hence the model is very predictive, the total number of parameters being nine in the Yukawa sector for the charged fermions and the neutrinos. 
In addition, it provides a dynamical origin for the mass hierarchy without invoking additional U(1) symmetry. Due to the ${ }^{(d)}T$ transformation property of the matter fields, the $b$-quark mass can be generated only when the ${ }^{(d)}T$ symmetry is broken, which naturally explains  the hierarchy between $m_{b}$ and $m_{t}$. 
The $Z_{12} \times Z_{12}^{\prime}$ symmetry, to a very high order, also forbids operators that lead to nucleon decays. In principle, a symmetry smaller than  $Z_{12} \times Z_{12}^{\prime}$ would suffice in getting realistic masses and mixing pattern; however, more operators will be allowed and the model would not be as predictive.  
We obtain the Georgi-Jarlskog relations for three generations. This inevitably requires non-vanishing mixing in the charged lepton sector, leading to correction to the tri-bimaximal mixing pattern. The model predicts non-vanishing $\theta_{13}$, which is related to the Cabibbo angle as, $\theta_{13}\sim \theta_{c}/3\sqrt{2}$.  In addition, it gives rise to a sum rule, $\tan^{2}\theta_{\odot} \simeq \tan^{2} \theta_{\odot, \mathrm{TBM}} - \frac{1}{2} \theta_{c} \cos\beta$,  which is a consequence of the Georgi-Jarlskog relations in the quark sector. This deviation could account for the difference between the experimental best fit value for the solar mixing angle and the value predicted by the tri-bimaximal mixing matrix.

\begin{acknowledgments}
We thank Carl Albright, Stefan Antusch and Chris Carone for useful comments.  
The work of M-CC was supported, in part, by the start-up funds from the University of California at Irvine. The work of KTM was supported, in part, by the Department of Energy under Grant no. DE-FG02-04ER41290. M-CC thanks the Particle Theory Group at Fermilab for its hospitality while this work was performed. 
\end{acknowledgments}






\end{document}